\documentclass[lnicst]{svmultln}
\usepackage{makeidx}  
\usepackage{graphics}
\usepackage{subfigure}
\usepackage{float}
\usepackage{amsmath}
\usepackage{amssymb}
\usepackage{graphicx}
\usepackage{setspace}

\floatstyle{ruled}
\newfloat{algorithm}{tbp}{loa}
\providecommand{\algorithmname}{Algorithm}
\floatname{algorithm}{\protect\algorithmname}
\begin{document}
\mainmatter              
\title{Social-aware Opportunistic Routing Protocol\\
based on User's Interactions and Interests}
\titlerunning{Social-aware Content-based Opportunistic Routing Protocol}  
%
\author{Waldir Moreira\inst{1} \and Paulo Mendes\inst{1} \and Susana
Sargento\inst{2}}
\authorrunning{Waldir Moreira et al.}   
%
\tocauthor{Waldir Moreira, Paulo Mendes, Susana Sargento}
\institute{SITILabs, University Lus{\'o}fona\\
\email{{waldir.junior, paulo.mendes}@ulusofona.pt}
\and
Instituto de Telecomunica\c{c}{\~o}es, University of Aveiro\\
\email{susana@ua.pt}}

\maketitle              

\begin{abstract}        
Nowadays, routing proposals must deal with a panoply of heterogeneous
devices, intermittent connectivity, and the users' constant need for
communication, even in rather challenging networking scenarios. Thus,
we propose a Social-aware Content-based Opportunistic Routing Protocol,
\emph{SCORP}, that considers the users' social interaction and their
interests to improve data delivery in urban, dense scenarios. Through
simulations, using synthetic mobility and human traces scenarios,
we compare the performance of our solution against other two social-aware
solutions, \emph{dLife} and \emph{Bubble Rap}, and the social-oblivious
\emph{Spray and Wait}, in order to show that the combination of social
awareness and content knowledge can be beneficial when disseminating
data in challenging networks.
\end{abstract}
\section{Introduction}

Given the advent of powerful mobile devices and the fast pace of today's
world, users crave connectivity while on the go. This leads to a networking
scenario with heterogeneous, mobile, and power-constraint devices,
as well as wireless networks with intermittent connectivity even in
urban scenarios, due to the presence of wireless shadowing, and the
existence of closed access points and expensive Internet services.
Moreover, users' requirements for ubiquitous data access is not aligned
with the current Internet architecture, since users are not interested
in knowing the location of data.

It has been shown that focusing on the content, rather then on the
host, we can improve the performance of challenged networks \cite{socialcast,contentplace}
by allowing an efficient direct communication between producers and
consumers of content. In addition, exploiting nodes' social interactions
and structure (i.e., communities \cite{bubble2011}, levels of social
interaction \cite{dlife,cipro}) has been shown efficient to increase
the performance of opportunistic routing. Thus, combining content
knowledge (i.e., content type, interested parties) with social proximity
shall bring benefits (faster, better content reachability) in challenged
networks.

\emph{SCORP} exploits social proximity and content knowledge to augment
the efficiency of data delivery in urban, dense scenarios. We show
the advantages that \emph{SCORP} brings to the operation of opportunistic
networks (in terms of delivery, cost and latency) through simulations
based on synthetic mobility and trace-based scenarios.

This paper is structured as follows. Section 2 briefly goes over the
related work. In Section 3, we present \emph{SCORP}. Section 4 presents
our evaluation study. In Section 5, conclusions and future work are
presented. It is worth noting that the words \emph{information}, \emph{data},
\emph{message}, and \emph{content} are used interchangeably throughout
this paper.

\section{Related Work\label{sec:Related-Work}}

Routing in opportunistic networks must be capable of dealing with
occasional contacts, intermittent connectivity, highly mobile nodes,
power and storage-constrained devices, and the possible nonexistence
of end-to-end paths. In the last couple of years, different social-aware
opportunistic routing solutions have emerged \cite{bookchapter} trying
to exploit the less volatile graph created by social proximity metrics
in relation to metrics reflecting the mobility behavior of nodes.

Now with content being introduced to social-aware opportunistic routing,
proposals can be classified as content-oblivious or content-oriented.
Among the social-aware content-oblivious proposals, \emph{Bubble}
\emph{Rap} \cite{bubble2011}, \emph{dLife} \cite{dlife}, and \emph{CiPRO}
\cite{cipro} are close in essence to \emph{SCORP}: all exploit social
proximity to devise forwarding schemes.

\textit{Bubble Rap} combines node centrality with the idea of community
structure to perform forwarding. Communities are formed considering
the number of contacts between nodes and their duration, and centrality
is seen from a local (i.e., inside communities) and global (i.e.,
whole network) perspective. Messages are replicated based on the global
centrality metric until it reaches the community of the destination
(i.e., a node belonging to the same community). At this point, forwarding
is done by using the local centrality metric, aiming to reach the
destination inside the community. 

With \emph{dLife} , the dynamism of users' behavior found in their
daily life routines is considered to aid routing. The goal is to keep
track of the different levels of social interactions (in terms of
contact duration) that nodes have throughout their daily activities
in order to infer how well socially connected users are in different
periods of the day. 

\emph{CiPRO} considers the time and place nodes meet throughout their
routines and\emph{ }holds knowledge of nodes (e.g., carrier's name,
address, nationality, device's battery level, memory) expressed by
means of profiles that are used to compute encounter probability among
nodes in specific time periods. 

While \emph{CiPRO} uses users' daily social interactions to classify
the type of contact among them, aiming to predict future encounters,
\emph{SCORP} uses these interactions to measure the proximity between
nodes sharing data interests. This is similar to what happens with
\emph{dLife} and \emph{Bubble} \emph{Rap}: the former weighs the levels
of social interaction between nodes and computes their importance;
and the latter uses social interactions to identify communities and
popular (i.e., high centrality) nodes.

Regarding the social-aware content-oriented proposals, \emph{SocialCast}
\cite{socialcast} and \emph{ContentPlace} \cite{contentplace} also
take into account the content and users' interest on it. 

\textit{SocialCast} considers the interest shared among nodes and
devises a utility function that captures the node's future co-location
(with others sharing the same interest) and the change in its connectivity
degree. Thus, the utility functions used by \emph{SocialCast} measure
how good message carrier a node can be regarding a given interest.
Moreover, \textit{SocialCast} functions are based on the publish-subscribe
paradigm, where users broadcast their interests, and content is disseminated
to interested parties and/or to high utility new carriers. Since the
performance of \emph{SocialCast} is related to the co-location assumption
(i.e., nodes with same interests spend quite some time together),
the proposal may be compromised in scenarios where it does not always
apply as such assumption may not always be true \cite{people}.

Besides taking into account the interest that users have in the content,\emph{
ContentPlace} \cite{contentplace} also considers information about
the users' social relationships to improve content availability. For
that, a utility function is computed for each data object considering
the access probability to the object and the involved cost in accessing
it, as well as the user's social strength towards the different communities
that he/she belongs to and/or has interacted with. The idea is to
have users fetching data objects that maximize the utility function
with respect to the local cache limitations, choosing the objects
that are of interest to him/herself and can be further disseminated
in the communities with which they have strong social ties.

These social-aware content-oriented approaches differ from \emph{SCORP}
as \emph{SocialCast} is based on the publish/subscribe paradigm (i.e.,
our solution does not require propagation of interests further than
encountered nodes), and \emph{ContentPlace} is much more data-aware:
besides the content type and interested parties, it also considers
how much content has already been spread and its availability. 

When making an overall analysis of all proposals, it is clear that
\emph{SCORP} may contribute to reduce network overhead and to make
routing rather simple when compared to \emph{SocialCast}, \emph{ContentPlace}
and \emph{CiPRO}, since it is independent from attributes such as:
i) connectivity degree and node co-location \cite{socialcast}; ii)
content availability, and users' communities \cite{contentplace};
iii) prediction of future encounter \cite{cipro}. Regarding content-oblivious
solutions such as \emph{dLife} and \emph{Bubble} \emph{Rap}, conceptually
it is not clear the advantages and limitations that content-oriented
proposals, such as \emph{SCORP,} may have in terms of the data dissemination
efficiency. Therefore, \emph{dLife} and \emph{Bubble} \emph{Rap} are
selected as benchmarks for our comparison studies. As we aim at a
low cost associated to message delivery, \emph{Spray and Wait }\cite{spraywait}
is considered as lower bound for delivery cost for being concerned
with resource usage (it controls replications to spare resources).
Hence, in a general sense, this paper aims to prove that taking content
into account leads to an improvement on the performance of social-aware
opportunistic routing, based on the performance of \emph{SCORP}. In
a future work, we aim to experimentally show the conceptually advantages
that \emph{SCORP} has in relation to \emph{SocialCast}, \emph{ContentPlace}
and \emph{CiPRO, }as soon as the code or a detailed specification
(e.g., Internet Draft) of such approaches is made available, to allow
us to perform a precise implementation, since details provided in
the papers are not enough to achieve such goal.

\section{The SCORP Proposal\label{sec:SCORP}}

This section presents our social-aware content-based opportunistic
routing proposal that takes into account the social proximity between
nodes and the content knowledge that nodes have while taking forwarding
decisions. \emph{SCORP} is based on a utility function that reflects
the \emph{probability of encountering nodes with a certain interest
among the ones that have similar daily social habits}. The reason
to use social proximity with content knowledge is two-fold: first,
nodes with similar daily habits have higher probability of having
similar (content) interest \cite{socialcast}; second, social proximity
metrics allow a faster dissemination of data, taking advantage of
the more frequent and longer contacts between closer nodes.

\begin{figure*}
\begin{centering}
\vspace{-0.4cm}
\includegraphics[scale=0.33]{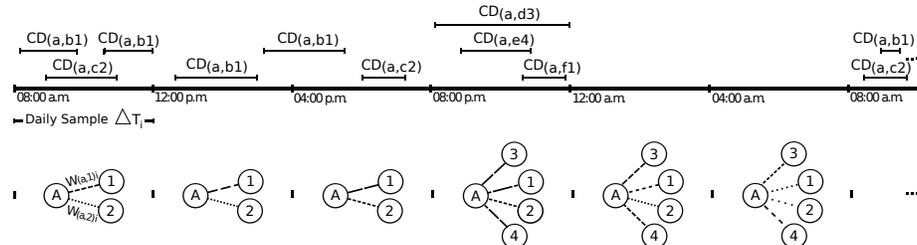}\vspace{-0.3cm}

\par\end{centering}

\caption{\label{fig:0}Contacts that node $A$ has with nodes having interests
$x$ ($CD(a,x)$) in different daily samples $\Delta T_{i}$.}
\vspace{-0.4cm}
\end{figure*}

Fig. \ref{fig:0} shows the different social interactions that a node
$A$ has with other nodes throughout its daily routine. For the sake
of simplicity, in this example each encountered node has only one
interest (nodes $B$ and $F$ have interest 1, and nodes $C$, $D$
and $E$ have interests $2$, $3$, and $4$, respectively). \emph{SCORP}
measures the duration of contacts, indexing such duration to interests
that such nodes have (cf. $CD(a,b1)$ in Fig. \ref{fig:0}). This
way, nodes have measures of different levels (intermittency of lines
in graphs) of social interactions with nodes having similar interests
($w(a,1)$) during specific time periods of their daily activities.
These different levels of social interactions are considered while
deciding whether a node is classified as a good forwarder for a message
tagged with a certain interest.

If a node $A$ has $n$ contacts with another node having an interest
$x$ in a daily sample $\Delta T_{i}$, with each contact $k$ having
a certain duration (\emph{Contact} \emph{Duration} - $CD\,(a,x)_{k}$),
at the end of $\Delta T_{i}$ the \emph{Total Connected Time to Interest
$x$} ($TCTI\,(a,x)_{i}$) is given by Eq. \ref{eq:1}.

\begin{equation}
TCTI\,(a,x)_{i}=\sum_{k=1}^{n}CD\,(a,x)_{k}\label{eq:1}
\end{equation}

The \emph{Total Connected Time to Interest} $x$ in the same daily
sample over consecutive days is used to estimate the average duration
of contacts towards the data interest $x$ for that specific daily
sample. Thus, from the perspective of node $A$, the \emph{Average
Total Connected Time to Interest} $x$ (\emph{ATCTI}) during a daily
sample $\Delta T_{i}$ in a day $j$ is given by a cumulative moving
average of $TCTI$ in that daily sample ($TCTI(a,x)_{ji}$), and the
$ATCTI$ during the same daily sample $\Delta T_{i}$ in the previous
day ($ATCTI(a,x)_{(j-1)i}$) as illustrated in Eq. \ref{eq:2}.

{\small 
\begin{equation}
ATCTI\,(a,x)_{ji}=\frac{{TCTI\,(a,x)_{ji}+(j-1)ATCTI\,(a,x)_{(j-1)i}}}{j}\label{eq:2}
\end{equation}
}{\small \par}

Then, node $A$ computes the \emph{Time-Evolving Contact to Interest
$x$ (TECI)} (cf. Eq. \ref{eq:3}) to determine its social strength
($w(a,x)_{i}$) towards nodes tagged with interest $x$ in a daily
sample $\Delta T_{i}$ based on the $ATCTI$ computed in that daily
sample and consecutive $t-1$ samples, where $t$ is the total number
of samples. In Eq. \ref{eq:3} $\frac{t}{t\text{+k-i}}$ represents
the time transitive property as in \emph{dLife} \cite{dlife}.

\begin{equation}
TECI=w\,(a,x)_{i}=\sum_{k=i}^{i+t-1}\frac{t}{t+k-i}ATCTI\,(a,x)_{k}\label{eq:3}
\end{equation}

\subsection{Algorithm}

The operation of \emph{SCORP }is very simple as illustrated in Alg.
\ref{alg:AlgorithmSCORP}: when the $CurrentNode$ meets a \emph{$Node_{i}$}
in a daily sample $\Delta T_{k}$, it gets a list with all content
interests \emph{$Node_{i}$} was faced with in that daily sample,
and the social weights towards the nodes having such interests (\emph{$Node_{i}$}.weightsToAllinterests
computed based on Eq. \ref{eq:3}). Additionally, \emph{$Node_{i}$}
sends a list of the messages it already carries (\emph{$Node_{i}$}.carriedMessages).
Then, every $Message_{j}$ in the buffer of the $CurrentNode$ is
replicated to \emph{$Node_{i}$ }if:
\begin{itemize}
\item \emph{$Node_{i}$} has interest (\emph{$Node_{i}$.}getInterests)
in the content of the message\linebreak{}
($Message_{j}$.getContentType ); or\\

\item \emph{The social weight of $Node_{i}$} towards a node having that
interest\linebreak{}
(i.e., $Message_{j}$.getContentType) is greater than the weight that
the\linebreak{}
$CurrentNode$ has towards any node with the same interest.
\end{itemize}
With this, \emph{SCORP} is expected to create replicas only to nodes
that indeed have interest in the content carried by the message to
be forwarded, or that have a strong relationship with nodes that have
that specific interest. Consequently, it is expected the creation
of less replicas improving resource usage and decreasing delivery
latency.

\begin{algorithm}
\caption{\label{alg:AlgorithmSCORP}Forwarding with \emph{SCORP}}

\textbf{\footnotesize begin}{\footnotesize \par}

{\footnotesize \hspace{0.1 cm}}\textbf{\footnotesize foreach}\emph{\footnotesize{}
$Node_{i}$ }{\footnotesize encountered by $CurrentNode$ }\textbf{\footnotesize do}{\footnotesize \par}

{\footnotesize \hspace{0.3 cm}receive(}\emph{\footnotesize $Node_{i}$}{\footnotesize .weightsToAllinterests
and }\emph{\footnotesize $Node_{i}$}{\footnotesize .carriedMessages)}\textbf{\footnotesize \vspace{-0.01 cm}}{\footnotesize \par}

{\footnotesize \hspace{0.3 cm}}\textbf{\footnotesize foreach}{\footnotesize{}
$Message_{j}$}\textbf{\footnotesize{} }{\footnotesize $\in$}\textbf{\footnotesize{}
}{\footnotesize buffer.($CurrentNode$) \& $\notin$ buffer(}\emph{\footnotesize $Node_{i}$}{\footnotesize )
}\textbf{\footnotesize do\vspace{-0.03 cm}}{\footnotesize \par}

{\footnotesize \hspace{0.6 cm}}\textbf{\footnotesize if}{\footnotesize{}
($Message_{j}$.getContentType $\in$}\emph{\footnotesize $Node_{i}$.}{\footnotesize getInterests)}\textbf{\footnotesize \vspace{-0.03 cm}}{\footnotesize \par}

{\footnotesize \hspace{0.9 cm}}\textbf{\footnotesize then}{\footnotesize{}
$CurrentNode$.replicateTo(}\emph{\footnotesize $Node_{i}$}{\footnotesize ,
$Message_{j}$)}\textbf{\footnotesize \vspace{-0.03 cm}}{\footnotesize \par}

{\footnotesize \hspace{0.6 cm}}\textbf{\footnotesize else if}{\footnotesize{}
(}\emph{\footnotesize $Node_{i}$.}{\footnotesize getWeightTo($Message_{j}$.getContentType)
$>$ }{\footnotesize \par}

{\footnotesize \hspace{2.5 cm}$CurrentNode$}\emph{\footnotesize .}{\footnotesize getWeightTo($Message_{j}$.getContentType)}\textbf{\footnotesize \vspace{-0.03 cm}}{\footnotesize \par}

{\footnotesize \hspace{0.9 cm}}\textbf{\footnotesize then}{\footnotesize{}
$CurrentNode$.replicateTo(}\emph{\footnotesize $Node_{i}$}{\footnotesize ,
$Message_{j}$)}\textbf{\footnotesize \vspace{-0.03 cm}}{\footnotesize \par}

\textbf{\footnotesize end}
\end{algorithm}

\section{Comparison Evaluation\label{sec:Comparison-Evaluation} }

\emph{SCORP} is evaluated against \emph{dLife} \cite{dlife,draftDlife},
a social-aware proposal based on users' daily life routines; \emph{Bubble
Rap} \cite{bubble2011}, a community-aware proposal; and \emph{Spray
and Wait} \cite{spraywait}, a social-oblivious solution that serves
as lower bound in what concerns delivery cost. This section starts
by presenting the used methodology and experimental settings, followed
by the results obtained based on synthetic mobility models and trace-based
scenarios. This section ends with a scalability analysis.

\subsection{Evaluation Methodology\label{sub:Evaluation-Methodology}}

The simulations are carried in the Opportunistic Network Environment
(ONE) simulator \cite{one}, considering the available implementations
of \emph{Spray and Wait, Bubble Rap }and\emph{ dLife }for this simulator. The code for \emph{SCORP}%
\footnote{http://siti.ulusofona.pt/aigaion/index.php/publications/show/406%
} is also available for reviewers to download and test it.

Results are presented with a 95\% confidence interval and are analyzed
in terms of average delivery probability (i.e., ratio between the
number of delivered messages and the total number of messages that
should have been delivered), average cost (i.e., number of replicas
per delivered message), and average latency (i.e., time elapsed between
message creation and delivery).

\subsection{Experimental Settings \label{sub:Settings}}

In our experiments we use two different mobility models: a synthetic
one and one based on human mobility traces. The synthetic mobility
model comprises different mobility patterns. It simulates a 12-day
interaction in the city of Helsinki between 150 nodes divided into
8 groups of people and 9 groups of vehicles. Each node has a 11-Mbps
WiFi interface with 100-meter communication range.

One vehicle group (10 nodes) follows the \emph{Shortest Path Map Based
Movement} mobility model and represents police patrols that randomly
choose destinations and use the shortest path to reach them: waiting
times rang from 100 to 300 seconds. The remaining 8 vehicle groups
(each with 2 nodes) represent buses following the \emph{Bus Movement}
mobility model with waiting times ranging from 10 to 30 seconds. The
speed of vehicles range from 7 to 10 m/s. 

The groups of people have different number of nodes: group A has 14
nodes; groups C, E, F, and G have 15 nodes each; groups B and D have
16 nodes each; and group H has 18 nodes. People have walking speeds
between 0.8 to 1.4 m/s following the \emph{Working Day Movement} mobility
model and may use the bus to move around. Each group was configured
to have different offices, meeting spots, and home locations. Each
person has an average of 8 daily working hours and walk around the
office with pause times between 1 minute and 4 hours. These people
also have a 50\% probability of having a leisure activity after work
which may be done alone or in group and last up to 2 hours.

The used CRAWDAD human traces \cite{cambridge-haggle-imote-content-2006-09-15}
including 36 nodes, for two months while Cambridge University students
moved throughout their daily routines. As general remark regarding
this dataset, the measurements that we did to prepare the configuration
of the experiments show that it has an average of 32 contacts per
hour among nodes and such contacts happen sporadically. Additionally,
the average number of formed community is approx. 6.7, where most
of them comprise almost all nodes.

The challenge faced to configure the experimentation set was related
to the different nature of the approaches being compared: although
\emph{Bubble Rap}, \emph{dLife} and \emph{SCORP} are social-aware
routing solutions, they differ in the sense that \emph{SCORP} is receiver-driven:
driven by interests that potential receivers have about specific content
traversing the network. The other two approaches, as well as \emph{Spray
and Wait,} are source-driven: driven by the need that a node has to
send data to a specific receiver. Hence, to provide a fair comparison,
and to show the potential of bringing the content knowledge into the
opportunistic routing realm, we put the four solutions under the same
load conditions. That is, the number of messages reaching the destinations
in each simulation is the same.

Thus, in the synthetic mobility scenario, a total of 6000 messages
are generated and expected to be received throughout the simulation
of \emph{Spray and Wait, Bubble Rap }and\emph{ dLife}. To achieve
the same number of messages to be received in \emph{SCORP}, 170 messages
with unique content are generated and each group of people has 10
different and randomly assigned interests that may or not overlap
fully or partially with the interests of other groups. By combining
the types of interests that are assigned to such groups and the number
of generated messages with content matching these interests, we end
up with 6000 messages to be delivered throughout \emph{SCORP}'s simulation.

In the human mobility trace scenario, with\emph{ Spray and Wait, Bubble
Rap }and\emph{ dLife} the source creates and sends 1, 5, 10, 20 and
35 different messages towards each destination. In the case of \emph{SCORP,
}the source creates 35 messages with different interests once, and
each receiver is configured with 1, 5, 10, 20, and 35 different interests.
Since node 0 is the source of these messages to the remaining 35 nodes,
this means that a total of 35, 175, 350, 700, and 1225 messages will
reach the destinations in any of the simulations done with \emph{Spray
and Wait, Bubble Rap},\emph{ dLife} and \emph{SCORP}. Nevertheless
the number of messages generated by the source is different for the
source- and receiver-driven approaches: for instance, in a configuration
with a \emph{dLife} or \emph{Bubble} \emph{Rap} source generating
20 different messages for each of the 35 nodes, we have a total of
700 messages being generated and expected to reach the destinations;
in the case of \emph{SCORP}, each of the 35 receivers is configured
with 20 different interests, so we have 35 messages being generated
and the same 700 messages are expected to reach the destinations.

The configurations of messages and interests (denoted in the paper
as msg/int in Sec. 4.4) are done to guarantee the same amount of potential
messages being delivered. The msg/int notation denotes the number
of different messages sent by \emph{Spray and Wait, Bubble Rap }and\emph{
dLife} sources or the number of different interests of each of the
\emph{SCORP} receivers.

Message TTL vary between 1, 2, 4 days, 1, and 3 weeks to represent
the different applications that cope with opportunistic networks,
and message size ranges from 1 to 100 kB. Although message TTL may
not be of great interest with the content-oriented paradigm if we
take into account that content can be always stored in the network,
we consider a more realistic scenario in which content utility is
timely limited. Hence, we chose to represent messages with different
TTL values. Message size ranges from 1 to 100 kB. Nodes have only
a 2 MB buffer space: despite the content-oriented concept consider
no buffer limitations as nodes are capable of storing large amount
of data, we assume that users may not be willing to share all the
storage capacity of their devices. Both message and buffer size follow
the universal evaluation framework proposed earlier \cite{latincom}.
To guarantee fairness throughout our comparison study for \emph{Spray
and Wait, Bubble Rap }and\emph{ dLife} in the human trace scenario,
node 0 has no buffer size restriction to avoid message discardation
due to buffer constraint given the number of messages it has to generate.
Additionally, the rate of message generation varies with the load:
when the load is of 1, 5, and 10 messages generated to each node,
they are generated at a rate of 35 messages per day. As for the load
with 20 and 35 messages, the rates are of 70, and 140 messages per
day, respectively. This is done to allow \emph{Bubble Rap} and \emph{dLife
}messages to be exchanged/delivered given the message TTL (i.e., 1
day).

We use the synthetic and the human traces mobility scenarios to analyze
different properties of the solutions being compared: the impact of
having different message TTLs in the case of the synthetic mobility
models; and the impact of having different network load in the case
of the human traces mobility models. We also observed the impact of
the different network load while varying the TTL, but these last set
of results have been omitted due to space limitation.

As for the proposals, \emph{Spray and Wait} runs in binary mode with
number of copies $L$ set to 10. \emph{Bubble Rap} uses algorithms
for community formation and node centrality computation (K-Clique
and cumulative window) \cite{bubble2011}. \emph{dLife }and \emph{SCORP}
consider 24 daily samples of one hour as mentioned in \emph{dLife}'s
paper \cite{dlife}.

\subsection{Evaluation of TTL Impact \label{sub:TTL-Impact-Evaluation}}

We use the synthetic mobility model with varying message TTL, in order
to: i) assess the impact that message TTL has on opportunistic routing
solutions; and ii) choose the TTL value that allows solutions to have
the best overall performance\emph{. }Before looking into the results,
here is a general remark regarding the synthetic mobility model: it
has an average of 962 contacts per hour happening in a homogeneous
manner.

\begin{figure}
\vspace{-0.4cm}
\subfigure[Average delivery probability]{\label{fig:1}
\includegraphics[scale=0.7]{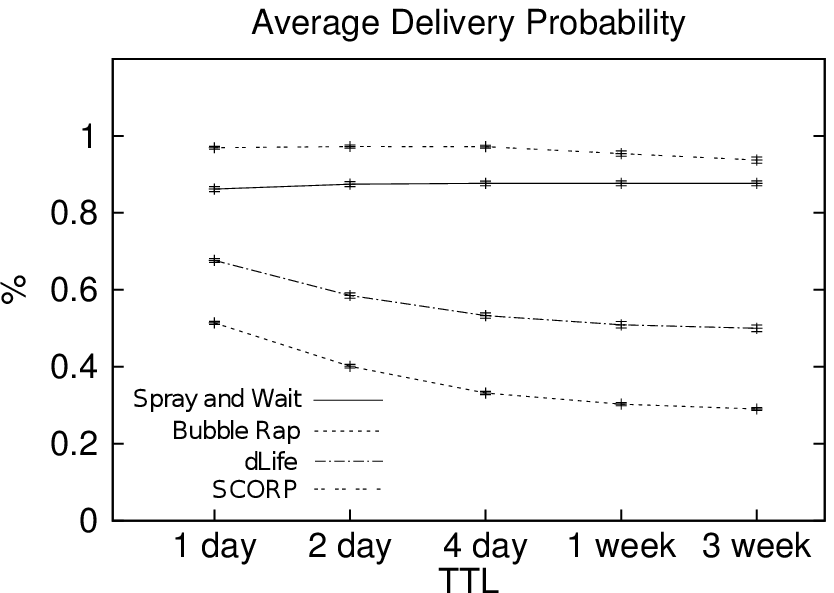}}~~~\subfigure[Average cost]{\label{fig:2}\includegraphics[scale=0.7]{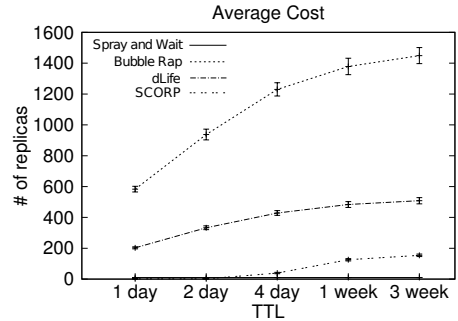}}
\subfigure[Average latency]{\label{fig:3}{\scriptsize \hspace{3.2  cm}}\includegraphics[scale=0.7]{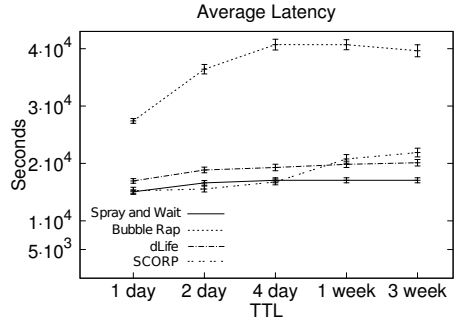}} 
\vspace{-0.3cm}
\caption{Performance under synthetic mobility model}
\vspace{-0.4cm}
\end{figure}

Fig. \ref{fig:1} shows the average delivery probability. The performance
of \emph{Bubble Rap} is affected by the fact that, while communities
are still being built it relies mostly on global centrality to reach
destinations. However, in this scenario, few nodes have high centrality
(20\%) and most messages are generated in low centrality nodes. As
a result replication is increased causing buffer exhaustion. This
situation gets worse as TTL increases.

\emph{dLife} performs up to 21\% better than \emph{Bubble Rap }as
it is able to capture the dynamic behavior of nodes. Given the high
number of contacts and their frequency, \emph{dLife} takes longer
to have a stable view of the network in terms of social weights, resulting
in useless replications leading to buffer exhaustion and preventing
more messages to be delivered.

\emph{Spray and Wait} outperforms \emph{Bubble Rap }and \emph{dLife}
(up to 58.6\% and 37.7\%, respectively): \emph{Spray and Wait} random
replications are able to reach most of these nodes, since the scenario
comprises buses and police patrols covering most of the simulated
area and equipped with a 100-meter transmission range. 

Since nodes interact very often, \emph{SCORP} also takes advantage
of shared interests among nodes to replicate content. Thus, messages
are quickly disseminated, increasing its delivery rate up to 64.7\%,
44.5\%, and 10.7\% over \emph{Bubble Rap}, \emph{dLife} and \emph{Spray
and Wait}, respectively. Still, \emph{SCORP} suffers a subtle decrease
of delivery rate due to the number of forwardings, which increases
with TTL. This causes few messages to be discarded due to buffer exhaustion,
since messages are allowed to live longer in the network.

When it comes to the average cost (cf. Fig. \ref{fig:2}), \emph{Bubble
Rap} creates the highest number of replicas to perform a successful
delivery, since it creates more replicas as messages are allowed to
live longer in the network \cite{bubble2011}. 

\emph{dLife} creates replicas based on the: i) social strength between
carrier/encountered nodes and destination; ii) node importance \cite{dlife}.
Since social weight is more accurate (i.e., capture reality) than
community formation given the subjective nature of the latter (communities
formed based on pre-defined and static contact duration, when in reality
people consider much more than this to create communities), this explains
why \emph{dLife} generates between 64.5\% and 65.2\% less replicas
than \emph{Bubble Rap} for the simulated TTLs.

Given the contact frequency in this scenario, \emph{SCORP }nodes have
a social weight towards all the different interests. This results
in a easier way to identify the nodes that should receive a replica
in order to successfully deliver content to interested nodes. As a
consequence \emph{SCORP} creates up to 99.8\% and 99.4\% less replicas
than \emph{Bubble Rap} and \emph{dLife}, respectively. 

\emph{Spray and Wait }serves as lower bound for delivery cost as it
limits the created number of replicas ($L$ = 10), thus it is expected
to have the best cost behavior (an average of 10.14 replicas across
the TTL configurations). Still, for the TTL configurations with one
and two days, \emph{SCORP} creates 8.6 and 8.3 less replicas when
compared to \emph{Spray and Wait.} This result show the advantage
in \emph{SCORP} for applications requiring low TTLs (messages with
a timely limited utility).

In terms of average latency (cf. Fig. \ref{fig:3}), \emph{Bubble
Rap} takes up to 58.1\%, 52.6\% and 58.8\% longer than \emph{Spray
and Wait},\emph{ dLife} and \emph{SCORP}, respectively, to deliver
content due to the fact that communities are not updated (i.e., nodes
not seen for a long period remain in communities), and few nodes have
high centrality. Thus, messages are replicated to nodes that have
weak social ties with the destination, which in turn increases the
overall time to deliver them.

\emph{dLife} and \emph{SCORP} experience less latency as forwardings
only happen when the encountered node: i) has higher social weight
towards the destination or is more important in the former case; and
ii) has a higher social weight towards a specific content (i.e., interest)
in the latter case, increasing their probability of delivering content
in less time. 

\emph{SCORP }has a subtle advantage over \emph{Spray and Wait }and
\emph{dLife }(up to 6.4\% and 17.6\% less latency, respectively) as
it considers the interest of nodes. This advantage is not seen for
TTL over 1 week: as messages are allowed to stay longer, \emph{SCORP}
takes more time to choose the best next forwarders.

We observe that the TTL has very little impact in social-oblivious
\emph{Spray and Wait}, while having an impact over the social-aware
proposals at different levels. Additionally, being content-oriented
has its advantage: \emph{SCORP} reaches a delivery rate of 97.2\%
with very little associated cost and low latency. 

This performance study led us to select the message TTL value that
allows the proposals to deliver the most messages in less time and
with the least associated cost. So, for the following set of results,
we use a 1-day message TTL.

\subsection{Evaluation of Network Load Impact \label{sub:Evaluation-of-LoadImpact}}

We use a human trace-based scenario with varying network load to assess
performance behavior of the studied proposals on a scenario with direct
exchange of data among mobile devices independently of the existing
levels of disruption/intermittency. As general remark: i) this dataset
has an average of 32 contacts per hour among nodes and contacts happen
sporadically; ii) with \emph{Bubble Rap} the average number of formed
community is approx. 6.7, where most of them comprise almost all nodes.

\begin{figure}
\vspace{-0.4cm}
\subfigure[Average delivery probability]{\label{fig:4}
\includegraphics[scale=0.7]{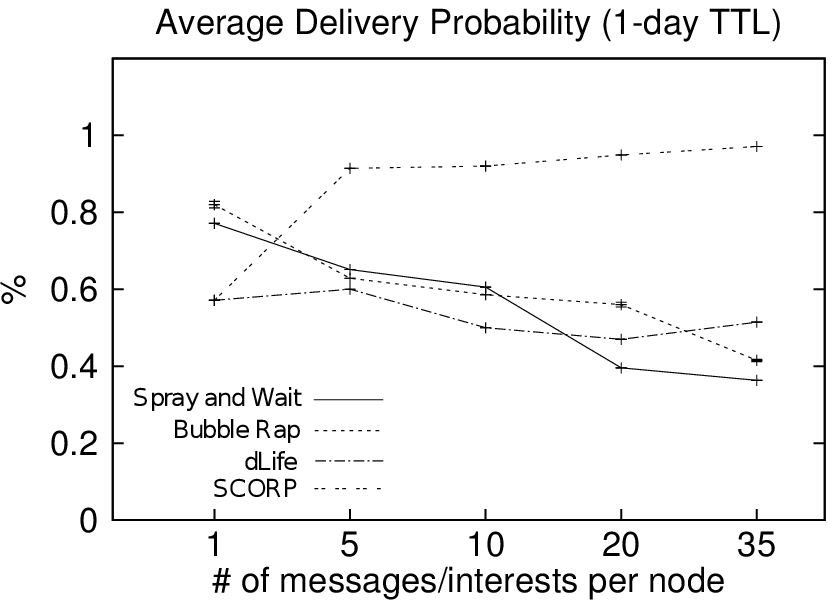}}~~~\subfigure[Average cost]{\label{fig:5}\includegraphics[scale=0.7]{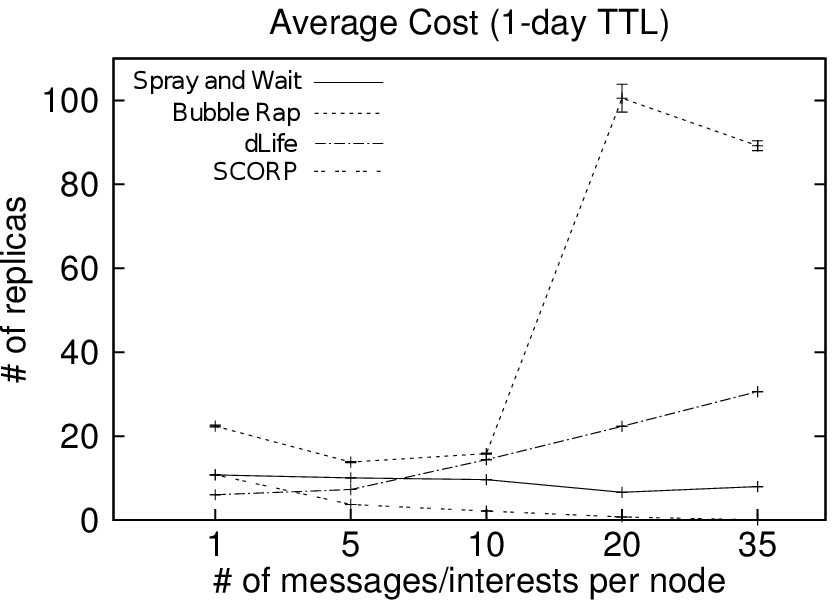}}
\subfigure[Average latency]{{\scriptsize \label{fig:6}\hspace{3.2  cm}}\includegraphics[scale=0.7]{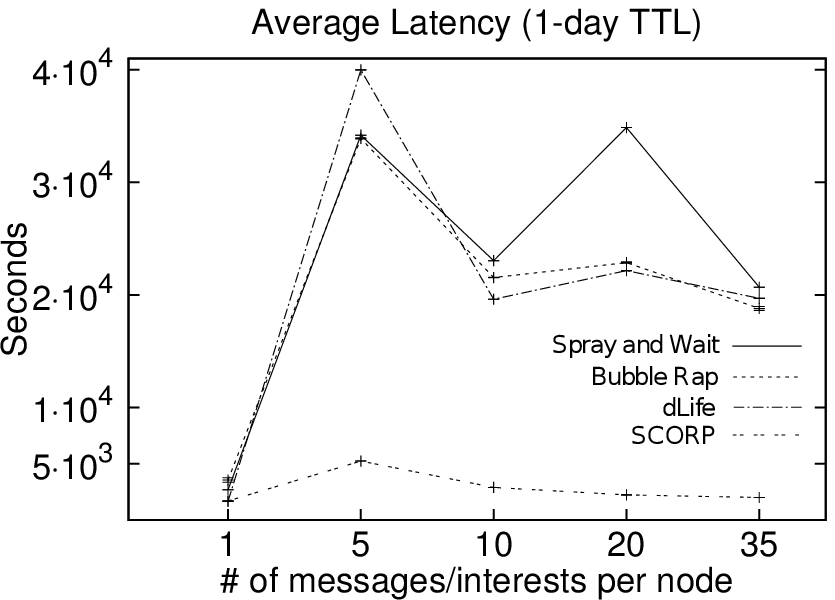}} 
\vspace{-0.3cm}
\caption{Performance under different network load}
\vspace{-0.4cm}
\end{figure}

Fig. \ref{fig:4} presents the results of average delivery probability
with an increasing number of messages/interests (msg/int) per node.
In the 1 msg/int configuration, \emph{Bubble Rap }delivers 4.9\% and
24.8\% more messages than \emph{Spray and Wait} and \emph{dLife}/\emph{SCORP},
respectively, since most of the communities comprise almost all nodes
and replication is done within those communities, resulting in more
replicas, and thus higher probability of delivering content. 

Despite of having a 20\% advantage over \emph{dLife} and \emph{SCORP}
regarding delivery, \emph{Spray and Wait }experiences a decrease in
performance when compared to the results described in Sec. 4.3. The
reason being that nodes in this scenario follow routines and do not
cover the whole simulated area. Consequently, replicas are created
to nodes that may never encounter the destination. 

\emph{dLife} and \emph{SCORP }have similar behavior, since forwarding
only occurs if social weight to nodes or node importance (\emph{dLife})
or social weight to interests (\emph{SCORP}) is greater in the encountered
nodes. Since contacts are little (32) and happen sporadically, these
proposals replicate less directly affecting their delivery capability.

For 5 and 10 msg/int configurations, the advantage of \emph{Spray
and Wait} and \emph{Bubble Rap} is reduced due to the limited TTL
and contact sporadicity: since messages can be created during a period
without contacts, they may never reach their destination.\emph{}

For \emph{Bubble Rap}, this issue is further increased in the 20 and
35 msg/int configurations, where it experiences buffer exhaustion.
We estimate buffer occupancy for the 20 msg/int configuration to support
this claim: there is an average of 39240 forwardings in the simulation,
if we divide this by the number of days (roughly 12%
\footnote{This dataset is worth of two months of data. However, when simulated
in ONE it is worth almost 12 day of communications. %
}) and by the number of nodes (35, source not included), we get an
average of 3270 replications per node. If we times this by the average
message size (52275 bytes), we get a buffer occupancy of 4.88 MB per
node, which exceeds the 2 MB allowed (cf. Sec. 4.1). This is just an
estimation for the worst case scenario with \emph{Bubble Rap} spreading
copies to every node. Since this is highly unlikely as\emph{ }it\emph{
}also uses centrality to control replication, buffer exhaustion worsens
as replication occurs to few nodes and not all as in our estimation.
As message generation rate increases with load, messages can potentially
take over forwarding opportunities of other messages, reducing the
delivery probability of the latter.

By considering the social strength towards destination or node importance,
\emph{dLife} has a stabler behavior when compared to \emph{Bubble
Rap}. Still, \emph{dLife} is affected by the rate of contacts due
to its design choices. We observe that buffer exhaustion (\textasciitilde{}24\%
more than the allowed) can also occur in this proposal in a 35 msg/int
configuration.

The performance of \emph{SCORP} shows the potential of content-awareness
in the context of opportunistic routing. The delivery ratio of \emph{SCORP}
increases as the ability of nodes to become a good message carrier
increases (i.e., the more interests a node has, the better it is to
deliver content to others since they potentially share interests). 

Regarding average cost (cf. Fig. \ref{fig:5}), \emph{Bubble Rap}
has the highest cost in the 1 msg/int configuration since it relies
on the formed communities to replicate: in an average of 671.4 forwardings
against the 317, 141 and 236 forwardings done by \emph{Spray and Wait},
\emph{dLife} and \emph{SCORP}, respectively. 

As expected, \emph{Spray and Wait }has a stable cost due to its limited
copies. In an attempt to find a next forwarder, nodes well socially
connected to the destination or to nodes interested in the content
carried in the message, \emph{dLife} and \emph{SCORP} solutions tend
to replicate less. 

\emph{SCORP} creates a few more replicas than \emph{dLife} due to
a particularity in its implementation: nodes interested in the content
of a certain message not only process that message, but also keep
a copy for further replication as they may have a chance to find nodes
with this same interest, or that met other nodes with such interest.
In this latter case, a node receiving a message with content matching
its interest, also replicates it (unnecessary and unwanted replicas)
to nodes that often have encountered it (and have a greater social
weight to that specific interest).\emph{}

For each of the 5, 10, 20 and 35 msg/int configurations, the number
of forwardings is proportional to the load. This is reflected in the
average cost of \emph{Bubble Rap }and \emph{dLife}: despite their
increased replication, their efforts are not enough to increase their
delivery rate and only contribute to the associated cost in delivering
content. 

With a greater list of interests, a \emph{SCORP} node can act~as
carrier for a larger number of nodes. Thus, those unwanted replicas
observed in the 1 msg/int configuration have a positive effect while
spreading content. Moreover, as messages are only replicated to interested
nodes or to nodes that have a stronger social weight towards other
nodes with higher interest in the content of the message than the
current carrier, the cost is reduced. Consequently, \emph{SCORP} creates
an average of \textasciitilde{}3.5 replicas across the msg/int configurations
against an average of 9, \textasciitilde{}48.4 and 16.1 replicas of
\emph{Spray and Wait},\emph{ Bubble Rap} and \emph{dLife}, respectively.
Moreover, \emph{SCORP} keeps resource usage (i.e., buffer) at a low
usage rate: with content awareness, the estimated maximum buffer occupancy
(as we have done for \emph{Bubble Rap}) varies between \textasciitilde{}0.03
MB (1 msg/int) and 0.15 MB (35 msg/int).

Fig. \ref{fig:6} shows the average latency that messages experienced.
For the 1 msg/int configuration, messages experience over 24\% and
52\% more latency with \emph{Bubble Rap} than with \emph{Spray and
Wait }and \emph{dLife}/\emph{SCORP}, respectively. This occurs since
nodes can be part of each others community and messages exchanged
between nodes take longer to reach destinations, due to the amount
of nodes within each community. 

By looking at the delivered messages, we observe that\emph{~dLife}
and \emph{SCORP} performed mostly (90\%) direct deliveries as the
source node meets destinations within the first two hours of simulation.
This surely reduces the overall latency, explaining why they take
the same time to perform a delivery. \emph{Spray and Wait} also delivered
most (85\%) of its messages by the second hour of simulation, but
only few (17\%) were directed delivered which reflects its random
replication power.

For the 5, 10, 20 and 35 msg/int configurations, we observe that the
latency peak experienced by all proposals is with the 5 msg/int configuration,
due to the message creation time. Messages are generated in a daily
basis and by analyzing the contacts/hour, we identified that some
messages are created during periods of very few (and sometimes no)
contacts followed by long periods (between 12 and 23 hours) of almost
no contact. Thus, messages are stored longer, contributing to the
overall high latency. This effect is mitigated as the load increases
(messages are created almost immediately before a high number of contacts),
reducing the experienced latency. As latency is in function of the
delivered messages, this explains the decrease and variable behavior
(from the 10 msg/int configuration on) experienced by\emph{ Spray
and Wait}, \emph{Bubble Rap }and \emph{dLife}: their delivery rate
decreases and increases, being influenced by the choices of next forwarders
that take longer to deliver content.\emph{}

\emph{SCORP} experiences up to 93.61\%, 90.25\% and 89.94\% less latency
than \emph{Spray and Wait}, \emph{Bubble Rap} and \emph{dLife}, respectively.
A \emph{SCORP} node can receive more information since it is interested
in the content being replicated, and becomes a better forwarder as
the chance of meeting nodes sharing the same interests is high. We
observe that almost all communities comprise almost all nodes. Although
the notion of community is not used in \emph{SCORP}, this observation
suggests that nodes have a high number of contacts, and this is advantageous
for \emph{SCORP,} as it can find interested nodes faster. To confirm
this claim, we look at the delivered messages, and observe that shared
interests account for 46\%, 53\%, 59\% and 66\% of deliveries in the
5, 10, 20, and 35 msg/int configurations, respectively. The remaining
destinations are reached by the ability that \emph{SCORP} has in identifying
interested parties further improving its performance.

\subsection{Scalability Analysis}

\emph{SCORP} takes into account users' interests in content. So, its
scalability is determined based on the total number of existing interests.
With this mind, we check memory requirements to compute~ $TECI$.
For a worst case scenario with $k$ time slots and $m$ interests,
and with every node meeting all other nodes (having at least one interest)
in each $\Delta Ti$, \emph{SCORP} requires: i) $m$ variables to
store every connection; ii) $m$ variables to store $TCTI$ computations;
and iii) $k\times m$ variables to store $ATCTI$ computations. Considering
each variable has $X$ bits, $TECI$'s needed resources is given by
Eq. \ref{eq:4}. 

\begin{equation}
TECI_{alloc}=m\times(k+2)\times X\, bits\label{eq:4}
\end{equation}

With 35 interests, 24 time slots, and 64 bit double for storing, \emph{SCORP}
requires 7.11 KB of storage in each node. However, content-driven
networks shall have a high number of interests: if per day a node
meets other nodes that have 1 billion different interests, \emph{SCORP}
requires 193.71 GB of memory, which is still feasible as nodes (e.g.,
laptops) have storage up to 500 GB. Still, not all nodes in dynamic
networks have such storage (i.e., smartphones) and even if they did,
owners would probably not share all of it on behalf of others. So,
a \emph{SCORP} node can reduce its encountered interest space by:
i) setting a daily threshold of 2 MB (equivalent to meet nodes with
more than 10000 interests); ii) eliminating the interests associated
to nodes not well socially connected to them at the end of a day:
iii)if the threshold is reached. This rules set the basics to allow
\emph{SCORP} to scale.

\section{Conclusions and Future Work\label{sec:Conclusions-and-Future}}

Access to data while on the go is desired by Internet users. Despite
of the available networking infrastructure, such goal can be rather
challenging, because most of the wireless access points is closed,
restricted or expensive, and wireless networks suffer from interference.

To overcome such challenges, an alternative is to allow direct exchange
of data among users by exploiting the type of content and the interest
users have on it \cite{socialcast,contentplace} along with social
similarly \cite{bubble2011,dlife,cipro} among users. This offloading
approach has shown its potential in improving data exchange over challenged
networking environments. 

Our study aims at further investigate the advantages of using the
content awareness (i.e., information type, interested parties) to
improve data dissemination in urban, dense scenario. Thus, we propose
\emph{SCORP,} a social-aware content-based opportunistic routing approach
based on users' daily interactions and interests. Our findings show
that the efficiency of data dissemination can be improved over challenged
networks when routing is designed having content knowledge and social
proximity in mind. \emph{SCORP} has better performance than previous
social-aware content-oblivious routing proposals (e.g,. \emph{Bubble
Rap} and \emph{dLife}): \emph{SCORP} delivers up to 97\% of its content
in an average of 46.9 minutes, against the 335.5 and 343.7 minutes
needed by \emph{Bubble Rap} and \emph{dLife}, respectively. Additionally,
\emph{SCORP} produces up to approximately 13.9 and 4.7 times less
replicas than \emph{Bubble Rap} and \emph{dLife}, respectively.

Since this work is part of the DTN-Amazon project%
\footnote{http://siti.ulusofona.pt/\textasciitilde{}dtnamazon/%
} that aims at promoting the social/digital inclusion of the riverside
communities in the northern region of Brazil, as future steps we will
implement \emph{SCORP} as content dissemination application among
the students of the Federal University of Para campus, Belem, Brazil,
and later being extended to disseminate content (public, health, safety)
to these isolated communities. Moreover, we would like to experimentally
show the conceptually advantages of \emph{SCORP} in relation to other
content-oriented social-aware solutions (\emph{SocialCast}, \emph{ContentPlace}
and \emph{CiPRO})\emph{ }as soon as the code of such approaches is
made available, or if guidance is available to support realistic,
unbiased implementations.

\section*{Acknowledgment}

Thanks are due to FCT for supporting the UCR (PTDC/EEA-TEL/103637/2008)
project and Mr. Moreira's PhD grant (SFRH/BD/62761/2009), and to the
colleagues of the DTN-Amazon project for the fruitful discussions.
%
%

\bibliographystyle{elsarticle-num}
\bibliography{bib-or}
\end{document}